\documentclass[twoside]{article}
\usepackage{amsmath,amsfonts,amssymb,latexsym,makeidx}
\usepackage{graphics,graphicx}
\usepackage{indentfirst,floatflt,cite}

\allowdisplaybreaks 

\begin{document}

\title{%
\bf Semiclassical symmetry of the Gross-Pitaevskii equation with
quadratic nonlocal Hamiltonian}

\author{Lisok $^{1}$\thanks{e-mail:lisok@phtd.tpu.edu.ru} A.L.,
Trifonov $^1$\thanks{e-mail: trifonov@phtd.tpu.edu.ru} A.Yu., and
Shapovalov $^2$\thanks{e-mail: shpv@phys.tsu.ru } A.V. }

\date{
$^1$ Mathematical Physics Department \\
Tomsk Polytechnic University,\\
         Lenin ave., 30, Tomsk, Russia, 634034 \\
$^2$ Theoretical Physics Department\\
Tomsk State University, \\
        Lenin ave., 36, Tomsk, Russia, 634050 \\
}

\maketitle

\begin{abstract}
The Cauchy problem for the Gross--Pitaevsky equation with quadratic
nonlocal nonlinearity is reduced to a similar problem for the
correspondent linear equation. The relation between symmetry
operators of the linear and nonlinear Gross--Pitaevsky equations is
considered.
\end{abstract}

\section*{Introduction}

Recent advances in formation of the Bose-Einstein condensates (BECs)
of alkaline metal vapors \cite{shapovalov:CORNELL,
shapovalov:PITAEVSKII-1, shapovalov:GROSS} have stimulated study of
the theoretical models describing behavior of nonlinear systems in
external fields. In the BEC models the local and nonlocal
Gross--Pitaevskii equations (GPEs) (which are  also called the
Hartree-type equations in the mathematical literature) are widely
used. Besides the BEC theory, the nonlocal GPE serves as a basic
equation in the models of quantum many-particle systems, nonlinear
optics, collective excitations in molecular chains, etc.

Let us write down the nonlocal  Gross--Pitaevskii equation as
\begin{eqnarray}
&\lbrace -i\hbar\partial _t +\hat {\mathcal
H}_{\varkappa}(t)\rbrace\Psi(\vec x,t)= \lbrace -i\hbar\partial _t
+\hat {\mathcal
H}(t)+\varkappa\hat V(t,\Psi (t))\rbrace\Psi(\vec x,t) =0, \label{shapovalov:GPE}\\
&\Psi(\vec x,t)\in L_2({\mathbb R}^n_x),\quad  \hat V(t,\Psi(t))=
\displaystyle\int\limits_{{\mathbb R}^n} d\vec y\,\Psi^*(\vec y,t)
V(\hat z,\hat w,t)\Psi(\vec y,t). \label{shapovalov:VTE-2}
\end{eqnarray}
Here the linear operators $\hat{\mathcal H}(t)={\mathcal H}(\hat z,
t)$ and $V(\hat z,\hat w,t)$ are the Weyl-ordered functions
\cite{shapovalov:KARASEVMASLOV} of time $t$ and of noncommuting
operators
\[ \hat z=( \hat{\vec p},\vec x)=(-i\hbar\partial/{\partial\vec x}, \vec x), \qquad
\hat w=(-i\hbar\partial/{\partial\vec y}, \vec y),
  \qquad \vec x,\vec y\in {\mathbb R}^n, \]
with commutators
\begin{equation}
[\hat z_k,\hat z_j]_-=[ \hat w_k,\hat w_j]_-=i\hbar J_{kj}, \quad
[\hat z_k,\hat w_j]_-=0,\qquad k,j=\overline{1,2n},
\label{shapovalov:COMM-1}
\end{equation}
$J =\|J_{kj}\|_{2n\times 2n}$ is a identity symplectic matrix
$ J=\left(\begin{array}{cc}0&-{\mathbb I}\\
{\mathbb I}& 0 \end{array} \right)_{2n\times 2n}, $ ${\mathbb
I}={\mathbb I}_{n\times n}$ is an identity $n\times n$-matrix.

In  multidimensional space GPE (\ref{shapovalov:GPE}) with variable
coefficients of general form is non-integrable by known methods
like, e.g., the Inverse Scattering Transform
\cite{shapovalov:ZAKHAROV-2}. Therefore, analytical solutions of
this equation can be constructed only approximately. An effective
approach to construct such solutions is provided by the method of
semiclassical asymptotics. Thus for nonlinear self-consistent field
operators, the theory of canonical operator with real phase was
constructed for solution of the Cauchy problem in \cite{Mas1,Mas2};
for spectral problems including singular potentials in \cite{Mas3,
Karas} (see also \cite{Bab, Molot, Vacul,Simenog}). Soliton-like
solutions of the Hartree-type equation with some potentials of
special form were constructed in \cite{Chetv}. A specific and
attractive feature of the nonlocal GPE with nonlinearity presented
in the equation only as a term under integral sign is that this
equation  can be referred to a class of nonlinear equations of
mathematical physics which are close to linear ones in a sense
\cite{shapovalov:KARASEVMASLOV}. Namely, among solutions there is a
subset of solutions regularly depending on the nonlinearity
parameter. Therefore, on the class of functions ${\mathcal P}_
\hbar^t$ called trajectory-concentrated functions
\cite{shapovalov:BTS1}, the problem of semiclassical asymptotics
construction for the nonlinear equation is reduced to an auxiliary
problem of construction of asymptotic solutions  for associated
linear  Schr\"odinger equations. Finally, the WKB-Maslov complex
germ method \cite{shapovalov:MASLOV-1,shapovalov:BEL-DOB} was
generalized for Eq. (\ref{shapovalov:GPE}). In particular, formal
solutions  of the Cauchy problem asymptotic in formal small
parameter $\hbar$ ($\hbar\to0$) was constructed accurate to
$O(\hbar^{N/2})$ where $N$ is any natural number. The leading term
of asymptotic solution of the spectral problem was found in
\cite{BLT07}.

Let us note that the semiclassical method, being approximate in
essence, allows one to find exact solutions in some special cases.
In \cite{Lisok:Sigma,shapovalov:LTS} the evolution operator for Eq.
(\ref{shapovalov:GPE}) with the quadratic nonlocal potential was
found. In this case the linear operators ${\mathcal H} (\hat z, t)$
and $ V(\hat z,\hat w,t)$ are quadratic in $\hat z$, $\hat w$:
\begin{eqnarray}
&&{\mathcal H} (\hat z, t)=\dfrac{1}{2}\langle\hat z,{\mathcal
H}_{zz}(t)\hat z\rangle+\langle{\mathcal H}_z(t),\hat z\rangle, \label{shapovalov:QUAD-1}\\
&& V(\hat z,\hat w,t)=\dfrac{1}{2}\langle\hat z,W_{zz}(t)\hat
z\rangle+\langle\hat z,W_{zw}(t)\hat
w\rangle+\dfrac{1}{2}\langle\hat w,W_{ww}(t)\hat w\rangle .
\label{shapovalov:QUAD-2}
\end{eqnarray}
Here, ${\mathcal H}_{zz}(t)$, $W_{zz}(t)$, $W_{zw}(t)$, $W_{ww}(t)$
are $2n\times 2n$-matrices, ${\mathcal H}_z(t)$ is a $2n$-vector;
$\langle .,.\rangle$ is an Euclidean scalar product of vectors:
$\langle\vec p,\vec x\rangle=\sum\limits^n_{j=1}p_jx_j$;\,\, $ \vec
p,\vec x\in {\mathbb R}^n$, $ \langle z,w\rangle=$
$\sum^{2n}_{j=1}z_jw_j$, $z,w\in {\mathbb R}^{2n}$.

The Hartree-type equations (\ref{shapovalov:GPE}),
(\ref{shapovalov:QUAD-1}), (\ref{shapovalov:QUAD-2}) are integrated
explicitly \cite{Lisok:Sigma,shapovalov:LTS} and possess fairly rich
symmetries, the study of which gives varied information about
solutions of the equation. It also shows how to reduce
multidimensional equations to one-dimensional equations, to
construct classes of exact and approximate solutions, to investigate
asymptotic of special classes of solutions, etc. Moreover, as long
as  Eqs. (\ref{shapovalov:GPE}), (\ref{shapovalov:QUAD-1}),
(\ref{shapovalov:QUAD-2}) have nonlocal nonlinearity, its symmetry
is of a special interest in the symmetry analysis. The matter is
that standard methods of symmetry analysis \cite{shapovalov:OVS,
shapovalov:IBRAGIM, shapovalov:OLVER, shapovalov:FUSCH-SS,
shapovalov:FUSCH-N}, developed basically for partial differential
equations (PDEs), face the problems when structure of equation under
the study differs from the  PDE because, for example, there are no
regular rules to choose an appropriate structure of symmetries for a
non-differential equation. Equation (\ref{shapovalov:GPE}) enable
one to avoid this problem as  symmetry of this equation  is closely
connected with the symmetry of linear equation corresponding to the
nonlinear one.

Here, we find symmetry operators of  Eqs. (\ref{shapovalov:GPE}),
(\ref{shapovalov:QUAD-1}), (\ref{shapovalov:QUAD-2})in explicit form
which, by definition, leave invariant the solution set of the
equation and  allow to generate  new solutions from the known ones,
(see, e.g., \cite{shapovalov:Miller, shapovalov:MANKO}).

\section{Operators in terms of an operator Caushy problem}

Consider the Cauchy problem for Eq. (\ref{shapovalov:GPE})
\begin{equation}
\Psi(\vec x,t,\hbar)|_{t=0}=\gamma(\vec x), \qquad \gamma\in
{\mathbb S},\qquad  \|\gamma(\vec x)||^2=1\label{shap2}
\end{equation}
where $\mathbb S$ is a Schwartz space.

The second-order Hamilton--Ehrenfest related to the Cauchy problem
(\ref{shapovalov:GPE}) (\ref{shap2}) has the form \cite{Lisok:Sigma}
\begin{equation}
\begin{cases}
 \dot z_{\Psi}=J\lbrace{\mathcal H}_z(t)+[{\mathcal H}_{zz}(t)+
\tilde\varkappa(W_{zz}(t)+W_{zw}(t))]z_{\Psi}\rbrace ,\label{shapovalov:HES}\\[8pt]
 \dot\Delta_{\Psi 2}=J[{\mathcal H}_{zz}(t)+\tilde\varkappa W_{zz}(t)]
 \Delta_{\Psi 2}-\Delta_{\Psi 2} [{\mathcal H}_{zz}(t)+\tilde\varkappa W_{zz}(t)]J.
\end{cases}
\end{equation}

Let us change the function $\Psi(\vec x,t)$ by $\Phi(\vec x,t)$ in
Eq. (\ref{shapovalov:GPE}) as
\begin{equation}
\Psi(\vec x,t)=\exp\Big\{\frac{i}\hbar[S(t)+\langle \vec P(t),\vec
x-\vec X(t) \rangle]\Big\}\Phi(\vec x-\vec X(t),t),\label{Zamena}
\end{equation}
where $S(t),\vec P(t),\vec X(t)$ are some differential functions to
be determined.

For $\Phi(\vec x,t)$ we have the following equation:
\begin{eqnarray}
& \Big\lbrace -i\hbar\partial_t +\dot S(t)+\langle \dot{\vec
{P}}(t),\vec x-\vec X(t) \rangle-\langle \vec P(t),\dot{\vec {X}}(t)
\rangle+\dfrac{1}{2}\langle\hat z_{\Phi},{\mathcal H}_{zz}(t)\hat
z_{\Phi}\rangle+ \langle {\mathcal H}_z(t),\hat
z_{\Phi}\rangle+\nonumber\\
& +\varkappa\displaystyle\int_{{\mathbb R}^n} d\vec y\,\Phi^*
\Big(\frac{1}{2}\langle\hat z_{\Phi},W_{zz}(t)\hat
z_{\Phi}\rangle+\langle\hat z_{\Phi},W_{zw}(t)\hat
w_{\Phi}\rangle+\frac{1}{2}\langle\hat w_{\Phi},W_{ww}(t)\hat
w_{\Phi}\rangle\Big) \Phi \Big\rbrace\Phi=0,\label{shapovalov:HES2}
\end{eqnarray}
where we use the notations $\hat
z_{\Phi}=(-i\hbar\partial/{\partial\vec x}+\vec P(t),\vec x)$; $\hat
w_{\Phi}=(-i\hbar\partial/{\partial\vec y}+\vec P(t),\vec y)$.

By changing variables $\vec x=\vec u+\vec X(t)$ and denoting
$\Phi(\vec u+\vec X(t),t)=\tilde\Phi(\vec u,t)$, we obtain the
following equation:
\begin{eqnarray}
&\bigg\lbrace -i\hbar\partial _t+i\hbar\langle \dot{\vec
{X}}(t),\dfrac\partial{\partial\vec u}\rangle +\dot S(t)+\langle
\dot{\vec {P}}(t),\vec u \rangle-\langle \vec P(t),\dot{\vec {X}}(t)
\rangle+\dfrac{1}{2}\langle\hat z_{\Phi u},{\mathcal H}_{zz}(t)\hat
z_{\Phi u}\rangle+\nonumber\\
&+\langle {\mathcal H}_z(t),\hat z_{\Phi u}\rangle
+\varkappa\displaystyle\int\limits_{{\mathbb R}^n} d\vec
y\,\tilde\Phi^*(\vec y ,t) \bigg(\frac{1}{2}\langle\hat z_{\Phi
u},W_{zz}(t)\hat z_{\Phi u}\rangle+\langle\hat z_{\Phi
u},W_{zw}(t)\hat w_{\Phi u}\rangle+\nonumber\\
&+\dfrac{1}{2}\langle\hat w_{\Phi u},W_{ww}(t)\hat w_{\Phi
u}\rangle\bigg) \tilde\Phi (\vec y ,t)\bigg\rbrace\tilde\Phi(\vec
u,t) =0, \label{Zamena10}
\end{eqnarray}
where $\hat z_{\Phi u}=\hat z_u+Z(t)$, $\hat w_{\Phi u}=\hat
z_y+Z(t)$,   $\hat z_u=(-i\hbar\partial/{\partial\vec u},\vec u)$,
$\hat z_y=( -i\hbar\partial/{\partial\vec y},\vec y)$.

The functions $\Phi(\vec u,t)$ are centered by construction, i.e.,
satisfy the condition
\begin{equation}
\int\limits_{{\mathbb R}^n} \tilde\Phi^*(\vec u,t)\hat z_u
\tilde\Phi(\vec u,t)d\vec u=0.
\end{equation}

Initial condition  (\ref{shap2}) for Eq. ({\ref{Zamena10}) has the
form
\begin{equation}
\tilde\Phi(\vec
u,t)\big|_{t=0}=\exp\Big\{-\frac{i}\hbar[S(0)+\langle \vec P(0),\vec
u\rangle]\Big\}\gamma(\vec u+\vec X(0)).\label{in_cond}
\end{equation}

Let the vector  $z=Z(t)=(\vec P(t),\vec X(t))$ satisfies the
equation
\begin{equation}
\dot Z(t)=J\lbrace{\mathcal H}_z(t)+[{\mathcal H}_{zz}(t)+
\tilde\varkappa(W_{zz}(t)+W_{zw}(t))]Z(t)\rbrace \label{Zamena2}
\end{equation}
with the initial condition  $Z(0)=\langle\gamma(\vec x)| \hat z|
\gamma(\vec x)\rangle$, and the function $S(t)$ is determined by the
relation
\begin{equation}
 S(t)= \int\limits_0^t \Bigl\lbrace\langle \vec P(t),\dot{\vec X}(t )\rangle
 -{\mathfrak H}(t)\Bigr\rbrace dt,\label{Zamena3}
\end{equation}
where
\begin{eqnarray*}
&&{\mathfrak H}(t)=\dfrac 12\langle Z(t),[{\mathcal H}_{zz}(t)+
\tilde\varkappa(W_{zz}(t)+2W_{zw}(t)+W_{ww}(t))]Z(t)\rangle+\\
&& \qquad +\langle {\mathcal H}_z(t),Z(t)\rangle+ \dfrac12
\tilde\varkappa \mbox{Sp}(W_{ww}(t)\Delta_{2}).
\end{eqnarray*}
Here the matrix $\Delta_{2}$ of the order $2n\times2n$ satisfies the
equation
\begin{equation}
\dot\Delta_{2}=J[{\mathcal H}_{zz}(t)+\tilde\varkappa W_{zz}(t)]
 \Delta_{2}-\Delta_{2} [{\mathcal H}_{zz}(t)+\tilde\varkappa
 W_{zz}(t)]J \label{Zamena4}
\end{equation}
and the initial condition
\begin{equation}
\Delta_{2}(0)=\frac{1}{2}\|\langle\gamma(\vec x)| \lbrace \Delta\hat
z_j\Delta\hat z_k+ \Delta \hat z_k\Delta\hat z_j\rbrace|\gamma(\vec
x)\rangle\|. \label{Zamena5}
\end{equation}

Then the function  $\tilde\Phi(\vec u,t)$ is a solution of the
linear associated equation
\begin{equation}
\Big\lbrace -i\hbar\partial_t+\frac{1}{2}\langle\hat
z_{u},({\mathcal H}_{zz}(t)+\varkappa W_{zz})\hat z_{u}\rangle
\Big\rbrace\tilde\Phi(\vec u,t)=0\label{LinUravnenie}
\end{equation}
with the initial condition \eqref{in_cond}.

Let $\hat A(\vec u,t)=A(\hat z_u,t)$ be the operator whose Weyl
symbol $A(z,t)$ satisfies the relation
\begin{equation}
\Big[ -i\hbar\partial_t +\frac{1}{2}\langle\hat z_{u},({\mathcal
H}_{zz}(t)+\varkappa W_{zz})\hat z_{u}\rangle ,\hat A(\vec u,t)\Big]
=0\label{Kommut}
\end{equation}
and the initial condition
\begin{equation}
 \widehat A(\vec u,t)|_{t=0}=\hat a(\vec u),
\end{equation}
where $\hat a(\vec u):{\mathbb S} \rightarrow{\mathbb S}$ is an
arbitrary operator.

Then, thereby (\ref{Kommut}) the operator  $\widehat A(\vec u,t)$ is
a symmetry operator of Eq. (\ref{LinUravnenie}) that maps a solution
$\tilde\Phi(\vec u,t)$ of Eq. (\ref{LinUravnenie}) into another
solution of this equation. Respectively, the function determined by
the condition
\begin{equation}
\overline{\Phi}_A(\vec u,t)=\frac{1}{\alpha_{A}}\widehat A(\vec
u,t)\tilde\Phi(\vec u,t),
\end{equation}
where  ${\alpha_{A}}=\|\hat a \tilde\Phi(\vec u,0)\|$ is also a
solution of Eq. (\ref{LinUravnenie}).

At $t=0$ we have
\begin{equation}
\overline\Phi_A(\vec u,0)=\frac{1}{\alpha_{A}}\hat a(\vec
u)\phi(\vec u) =\phi_A(\vec u).
\end{equation}
Here $||\phi_A(\vec u)||=1$, and we immediately obtain
$||\tilde\Phi_A(\vec u,t)||=1$. However, in general case the
function $\tilde\Phi_A(\vec u,t)$ can correspond no solution of the
original nonlinear equation as it is centered not for all $\widehat
A(\vec u,t)$:
\begin{equation}
\int\limits_{{\mathbb R}^n} \overline\Phi^*_A(\vec u,0)\hat z_u
\overline\Phi_A(\vec u,0)d\vec u\ne0.
\end{equation}

To find solutions of the original nonlinear equation which would
correspond to  $\tilde\Phi(\vec u,t)$, let us introduce the
notations
\begin{equation}
\lambda_0=\int\limits_{{\mathbb R}^n} \overline{\Phi}^*_A(\vec
u,0)\hat z_u \overline{\Phi}_A(\vec u,0)d\vec
u=\int\limits_{{\mathbb R}^n}\phi^*_A(\vec u) \hat z_u\phi_A(\vec
u)d\vec u
\end{equation}
and
\begin{equation}
\lambda(t)=\int\limits_{{\mathbb R}^n} \overline\Phi^*_A(\vec
u,t)\hat z_u \overline\Phi_A(\vec u,t)d\vec u.
\end{equation}
We can verify immediately that if  $\lambda(t)$ is  a solution of
the Cauchy problem
\begin{equation}
\dot{\lambda}(t)=J({\mathcal H}_{zz}(t)+\tilde\varkappa
W_{zz}(t))\lambda(t),\qquad \lambda(t)=\begin{pmatrix}
{\vec\lambda}_p(t) \\ {\vec\lambda}_u (t)\end{pmatrix}, \quad
\lambda(0)=\lambda_0=\begin{pmatrix} \vec\lambda_{p_0} \\
\vec\lambda_{u_0}\end{pmatrix},\label{shapovalov:MATRICIANT-100}
\end{equation}
then the function
\begin{equation}
 \tilde\Phi_A(\vec u,t)=\exp\Big\{\frac{i}\hbar[S_{\lambda}(t)+\langle \vec
\lambda_p(t),\vec u+\vec \lambda_u (t) \rangle]\Big\}
\overline{\Phi}_A(\vec u+\vec \lambda_u (t),t),
\end{equation}
where
\begin{equation}
S_{\lambda}(t)=\int\limits_0^t \Bigl\lbrace\langle \vec
\lambda_p(t),\dot{\vec \lambda}_u (t )\rangle - {\mathfrak
H}_{\lambda}(t)\Bigr\rbrace dt
\end{equation}
and
\[ {\mathfrak H}_{\lambda}(t)=\frac12\langle \lambda(t),[{\mathcal H}_{zz}(t)+
\tilde\varkappa(W_{zz}(t)+2W_{zw}(t)+W_{ww}(t))]\lambda(t)\rangle +
  \frac 12\tilde\varkappa \mbox{Sp}(W_{ww}(t)\Delta_{2}) \]
is a solution of Eq. (\ref{LinUravnenie}) and satisfies  the
condition
\begin{equation}
\int\limits_{{\mathbb R}^n} \tilde\Phi^*_A(\vec u,t)\hat z_u
\tilde\Phi_A(\vec u,t)d\vec u=0.
\end{equation}

Let us correlate according to (\ref{Zamena}) the functions
$\tilde\Phi_A(\vec u,t)=\Phi(\vec x + \vec X_A(t),t)$ and
$\tilde\Phi(\vec u,t)=\Phi(\vec x+\vec X(t),t)$ with the functions
$\Psi_A(\vec x,t)$ and $\Psi(\vec x,t)$, respectively. Then we have
the relation
\begin{eqnarray}
& \exp\Big\{-\dfrac{i}\hbar[S_A(t)+\langle \vec P_A(t),\vec x-\vec
X_A(t) \rangle]\Big\}\Psi_A(\vec
x,t)=\nonumber\\
& =\dfrac{1}{\alpha_{A}}\exp\Big\{\dfrac{i}\hbar
[S_{\lambda}(t)+\langle \vec \lambda_p(t),\vec x-\vec X_A(t)+\vec
\lambda_u (t) \rangle]\Big\}A(\vec x-\vec
X_A(t)+\lambda_u(t),t)\times\nonumber\\
& \times \exp\Big\{-\dfrac{i}\hbar[S(t)+\langle \vec P(t),\vec
x-\vec X(t) \rangle]\Big\}\Psi(\vec x+\vec X(t)-\vec
X_A(t)+\vec\lambda_u (t),t)
\end{eqnarray}
from which it follows
\begin{eqnarray}
& \Psi_A(\vec x,t)=\exp\Big\{\dfrac{i}\hbar[S_A(t)+\langle \vec
P_A(t),\vec x-\vec X_A(t) \rangle]\Big\}\times\nonumber\\
& \times\dfrac{1}{\alpha_{A}} \exp\Big\{\dfrac{i}\hbar
[S_{\lambda}(t)+\langle \vec \lambda_p(t),\vec x-\vec X_A(t)+\vec
\lambda_u (t) \rangle]\Big\} A(\vec x-\vec X_A(t)+\lambda_u(t),t)\times\nonumber\\
& \times \exp\Big\{-\dfrac{i}\hbar [S(t)+\langle \vec P(t),\vec
x-\vec X(t) \rangle]\Big\}\Psi(\vec x+\vec X(t)-\vec
X_A(t)+\vec\lambda_u (t),t).\label{lisok:eq5}
\end{eqnarray}
The operator
\[ \Psi_A(\vec x,t)=\widehat A_{\text{nl}}\Psi(\vec x,t) \]
determined by (\ref{lisok:eq5}) is a symmetry operator of the
original nonlinear equation.

\section{Symmetry operators obtained via auxiliary  equation }

Let  $\hat a:S \rightarrow S$ be an operator and $\gamma(\vec x) \in
S $ is a function. Let us set for Eq. (\ref{shapovalov:GPE}) the
following Cauchy problems: $\Psi(\vec x,t,\hbar)|_{t=0}=\gamma(\vec
x)$, $\gamma(\vec x)\in {\mathcal P}^0_\hbar$, $\|\gamma(\vec
x)||^2=1$, and $\Psi_A(\vec x,t,\hbar)|_{t=0}=\hat a\gamma(\vec
x)=\gamma_A(\vec x)$, $\gamma_A(\vec x)\in {\mathcal P}^0_\hbar$,
$\|\gamma_A(\vec x)||^2=1$.

Let us change the functions $\Psi(\vec x,t)$ and $\Psi_A(\vec x,t)$
in Eq. (\ref{shapovalov:GPE}) by the functions $\tilde\Phi(\vec
u,t)$ and $\tilde\Phi_A(\vec u,t)$, respectively, according to
formulas (\ref{Zamena}), (\ref{Zamena2}), (\ref{Zamena3}), and
(\ref{Zamena4}). Then the function $\tilde\Phi(\vec u,t)$ satisfies
the equation
\begin{equation}
\Big\{-i\hbar\partial_t +\frac{1}{2}\langle\hat z_{u},({\mathcal
H}_{zz}(t)+\varkappa W_{zz})\hat z_{u}\rangle\Big\}\tilde\Phi
=0\label{lisok1}
\end{equation}
and the initial condition $\tilde\Phi(\vec
u,t)\big|_{t=0}=\exp\Big\{-\dfrac{i}\hbar[S(0)+\langle \vec
P(0),\vec u\rangle]\Big\}\gamma(\vec u+\vec X(0))$, where the vector
$Z(t)$ is a solution of Eq. (\ref{Zamena2}) with the initial
condition $Z(0)=\langle\gamma(\vec x)|\hat z|\gamma(\vec x)\rangle$.
Similarly,  $\tilde\Phi_A(\vec u,t)$ is a  solution of the equation
\begin{equation}
\Big\{-i\hbar\partial_t +\frac{1}{2}\langle\hat z_{u},({\mathcal
H}_{zz}(t)+\varkappa W_{zz})\hat z_{ u}\rangle
\Big\}\tilde\Phi_A(\vec u,t) =0\label{lisok1a}
\end{equation}
with the initial condition $\tilde\Phi_A(\vec
u,t)\big|_{t=0}=\exp\Big\{-\dfrac{i}\hbar[S_A(0)+\langle \vec
P_A(0),\vec u\rangle]\Big\}\gamma_A(\vec u+\vec X_A(0))$.

Vector $Z_A(t)$ satisfies Eq. (\ref{Zamena2}) with the initial
condition $Z_A(0)=\langle\gamma_A(\vec x)|\hat z|\gamma_A(\vec x)
\rangle$, and
\begin{equation}
S_A(t)=\int\limits_0^t \Bigl\{\langle \vec P_A(t ),\dot{\vec
X}_A(t)\rangle-{\mathfrak H}_A(t)\Bigr\}dt,
\label{shapovalov:EVOLUT2}
\end{equation}
where
\begin{eqnarray}
&{\mathfrak H}_A(t)=\dfrac 12\langle Z_A(t),[{\mathcal H}_{zz}(t)+
\tilde\varkappa(W_{zz}(t)+2W_{zw}(t)+W_{ww}(t))]Z_A(t)\rangle
+\nonumber\\
&+\langle {\mathcal H}_z(t),Z_A(t)\rangle+ \dfrac12\tilde\varkappa\,
\mbox{Sp}(W_{ww}(t)\Delta_{2A}).
\end{eqnarray}
Matrix  $\Delta_{2A}(t)$ is a solution of Eq. (\ref{Zamena4}) with
the initial condition
\begin{equation}
\Delta_{2A}(0)=\frac{1}{2}\|\langle\gamma_A(\vec x,t)| \lbrace
\Delta\hat z_j\Delta\hat z_k+ \Delta \hat z_k\Delta\hat
z_j\rbrace|\gamma_A(\vec x,t)\rangle\|. \label{shapovalov:SIGMA}
\end{equation}
The relation  $\gamma_A(\vec x)=\hat a\gamma(\vec x)$  determines
operator $\hat{\bar{a}}(\vec u)$ such that $\tilde\Phi_A(\vec
u,0)=\hat{\bar{a}}(\vec u)\tilde\Phi(\vec u,0)$.

Consider the Cauchy problem for an operator  $\widehat{\overline
A}(\vec u,t)=\overline A(\hat z_u,t)$:
\begin{equation}
\Big[ -i\hbar\partial _t +\frac{1}{2}\langle\hat z_{u},({\mathcal
H}_{zz}(t)+\varkappa W_{zz})\hat z_{u}\rangle ,\hat A(\vec u,t)\Big]
=0\label{lisokA}
\end{equation}
with the initial condition
\begin{equation}
 \widehat{\overline A}(\vec u,t)|_{t=0}=\hat {\bar{a}}(\vec u).
\end{equation}
In accordance with (\ref{lisokA}) $\widehat A(\vec u,t)$ is a
symmetry operator of  Eq.  (\ref{lisok1}) and it maps solution
$\tilde\Phi(\vec u,t)$ of Eq.  (\ref{lisok1}) into its solution
$\tilde\Phi_A(\vec u,t)$.

If we correlate the functions  $\tilde\Phi_A(\vec u,t)=\Phi(\vec x +
\vec X_A(t),t)$ and  $\tilde\Phi(\vec u,t)=\Phi(\vec x+\vec X(t),t)$
with the functions  $\Psi_A(\vec x,t)$ and $\Psi(\vec x,t)$,
respectively, according to formula (\ref{Zamena}), we obtain
\begin{eqnarray}
& \exp\Big\{-\dfrac{i}\hbar[S_A(t)+\langle\vec P_A(t),\vec x-\vec
X_A(t) \rangle]\Big\}\Psi_A(\vec x,t)=\widehat{\overline{A}}(\vec
x-\vec X_A(t),t)\times\nonumber\\
& \times\exp\Big\{-\dfrac{i}\hbar[S(t)+\langle \vec P(t),\vec x-\vec
X(t)\rangle]\Big\}\Psi(\vec x+\vec X(t)-\vec X_A(t),t)
\end{eqnarray}
or
\begin{eqnarray}
& \Psi_A(\vec x,t)=\exp\Big\{\dfrac{i}\hbar[S_A(t)+\langle \vec
P_A(t),\vec x-\vec X_A(t) \rangle]\Big\} \widehat{\overline{A}}(\vec
x-\vec X_A(t),t)\times\nonumber\\
& \times\exp\Big\{-\dfrac{i}\hbar[S(t)+\langle \vec P(t),\vec x-\vec
X(t)\rangle]\Big\}\Psi(\vec x+\vec X(t)-\vec
X_A(t),t).\label{lisok:eq01}
\end{eqnarray}

An operator
\[ \Psi_A(\vec x,t)=\widehat{\overline{A}}_{\text{nl}}\Psi(\vec x,t) \]
determined by relation \eqref{lisok:eq01} is also a symmetry
operator of the initial nonlinear equation.

\section{Conclusion}

It should be noted that a direct calculation of symmetry operators
for a nonlinear equation is, as a rule,  difficult because of the
complexity of the determining equations
\cite{shapovalov:PUKHNACHEV}. In construction of symmetry operators, we have used the fact that
the original nonlinear equation can be connected with an associated
linear equation, and for quadratic operators of the form
(\ref{shapovalov:QUAD-1}) and (\ref{shapovalov:QUAD-2}) these
associated equations coincide. In general, this is not true and
therefore, analytical solutions of the GPE can be constructed only
approximately.

Many approximate methods are based on appropriate anzats
representing the general element of a class of functions in which
approximate solution is constructed. To such methods it is possible
to relate, e.g., the  method of collective variables (see, e.g.,
\cite{col1} and references therein), the Lagrangian method
\cite{lag1, lag2}, etc.

It appears that the most effective method in various
multidimensional problems of mathematical physics is the method of
semiclassical approximations. Peculiarity that differs this method
it from the usual method of expanding in power series in a small
asymptotic parameter is that the asymptotic small parameter is
included into the solution both regularly and singularly. This fact
allows us, in particular, to construct space-localized solutions
having  an important physical meaning. This possibility is of a
special importance in nonlinear problems where stable localized
excitations (patterns), such as, for example, solitons are the
object of research. The advantage of the semiclassical asymptotic
method  is that it allows to estimate accuracy of the constructed
solution with a given power of the asymptotic parameter.

Symmetry analysis provides another approach to construct  analytical
solutions  of the nonlinear equations \cite{shapovalov:OVS,
shapovalov:OLVER, shapovalov:FUSCH-SS, shapovalov:IBRAGIM,
shapovalov:FUSCH-N}. Taking into account invariant properties of the
equation under consideration, one can find classes of special
solutions which can serve as a prototype of anzatses for classes of
exact solutions.

However, variable coefficients, as a rule, reduce the equation
symmetry and even can exclude it completely that reduces the
symmetry analysis in such cases. Comparison of these two  approaches
results in a problem of consideration of approximate symmetry for
the GPE in the formalism of semiclassical approximation method (see,
e.g., \cite{shapovalov:SHVEDOV}).

\subsection*{Acknowledgements}
The work was supported in part by  President of the Russian
Federation, Grant  No SS-5103.2006.2, MK-4629.2006.2, DFG.


\end{document}